# Kagome van-der-Waals $Pd_3P_2S_8$ with flat band

Seunghyun Park[1,2], Soonmin Kang[1], Haeri Kim[1], Ki Hoon Lee[1,2], Pilkwang Kim[3], Sangwoo Sim[1,3], Nahyun Lee[1], Balamurugan Karuppannan[1], Junghyun Kim[1], Jonghyeon Kim[4], Kyung Ik Sim[4], Matthew J. Coak[1], Yukio Noda[5], Cheol-Hwan Park[3], Jae Hoon Kim[4], and Je-Geun Park[1,3,6*]

[1]Center for Correlated Electron Systems, Institute for Basic Science (IBS-CCES), Seoul, 08826, Republic of Korea
[2]Department of Physics, Incheon National University, Incheon 22012, Republic of Korea
[3]Department of Physics and Astronomy, Seoul National University, Seoul, 08826, Republic of Korea
[4]Department of Physics, Yonsei University, Seoul, 03722, Republic of Korea
[5]Institute of Multidisciplinary Research for Advanced Materials, Tohoku University, Sendai, 980-8577, Japan
[6]Center for Quantum Materials, Seoul National University, Seoul, 08826, Republic of Korea

[*]jgpark10@snu.ac.kr

## Abstract

With the advanced investigations into low-dimensional systems, it has become essential to find materials having interesting lattices that can be exfoliated down to monolayer. One particular important structure is a kagome lattice with its potentially diverse and vibrant physics. We report a van-der-Waals kagome lattice material, $Pd_3P_2S_8$, with several unique properties such as an intriguing flat band. The flat band is shown to arise from a possible compact-localized state of all five *4d* orbitals of Pd. The diamagnetic susceptibility is precisely measured to support the calculated susceptibility obtained from the band structure. We further demonstrate that $Pd_3P_2S_8$ can be exfoliated down to monolayer, which ultimately will allow the possible control of the localized states in this two-dimensional kagome lattice using the electric field gating.

Introduction

Over the past two decades, two-dimensional (2D) van-der-Waals (vdW) materials have played an active role in the study of condensed matter physics[1,2]. Reducing the spatial dimension not only allows the realization of the well-known two-dimensional models like Ising, XY, and Heisenberg models but also is expected to manifest novel quantum phenomena as in twisted bilayer graphene.[3] More recent studies of magnetic vdW systems[4-7]: in particular, $TMPS_3$ (TM=Fe, Ni, Mn)[8-10], $CrI_3$[11], $Fe_3GeTe_2$[12], and $CrGeTe_3$[13] have provided an exciting route to understanding the spin-related nanophysics. Nonetheless, exploring the 2D magnetic properties is still restricted to 2D materials, typically having honeycomb lattice. Therefore, introducing a new lattice into the still relatively limited collection of available materials will be extremely beneficial for the 2D magnetic materials community and related fields.

Kagome crystal structure is exceptionally rare among the known vdW materials and yet is related to subjects in other diverse fields of condensed matter physics such as topological insulators[14], the fractional quantum Hall effect[15-17], and proposed quantum spin liquids[18-21]. The exotic properties of the kagome lattice originate from their unique structure, a strongly spin-frustrated two-dimensional structure of corner-sharing triangles, as shown in Figure 1a. Another unique feature originating from this frustrated system is a completely flat band, as in Figure 1b. The flat band naturally arises from the destructive interference of electron wave functions. It thus represents the localized electrons of compact localized states (CLSs) in real space, as shown in Figure 1a. A set of CLSs inducing the flat band may superpose with one another without costing any additional energy. The localization of electrons, therefore, is naturally expected to increase the correlation effects[22-24]. Such correlation effects are expected to give rise to unusual phenomena such as fractional quantum Hall effect[15], ferromagnetism[25,26], and high $T_c$ superconductivity[18,27-28], to name only a few. Once we have a kagome van-der-Waals material that can be cleaved down to monolayer, it will provide an ideal condition for exploring those physics on two dimensions[24-26].

The lack of adequate materials having the kagome lattice has been a bottleneck in the related field despite those many fascinating theoretical predictions. Most of the experimental realizations have so far been limited to artificial kagome or bulk kagome systems, which contain 2D kagome embedded in a three-dimensional crystal[29-36]. Indeed, studies on the bulk kagome materials, including metallic kagome[31-35], Herbertsmithite[18], and other metal-organic frameworks[36] have been the playground for the exploration into several intriguing electric and magnetic properties of the frustrated systems and topological states. However, the interlayer interaction of these bulk materials, although weak, is likely to mask the very nature of kagome physics and hamper its emergent phenomena[31]. A recent discovery of bulk kagome materials such as FeSn and CoSn also provides an opportunity to examine the kagome properties using bulk samples[34,35]. That being said, obtaining a monolayer kagome remains a considerable challenge and, at the same time, could offer enormous potential, if successful. With an isolated kagome layer, one can imagine building nanostructures with a kagome lattice material to study the physical phenomena such as Hall effects, layer-dependent, and twisted-angle-dependent electronic structures[3,32,37].

We report $Pd_3P_2S_8$ as a promising candidate to explore ideal 2D kagome physics. The material can produce a monolayer with the band structures having a very unusual electronic structure of a distinctive flat band[38-40]: an exciting topic, as discussed in other studies[15-17, 22-25]. Our magnetization measurements are in good agreement with the susceptibility calculated from the band structure. We discuss how the unique flat-band can arise from the frustrated kagome structure and show that the flat band is robust down to the monolayer.

## Results

We prepared the samples following the recipes summarized in Method (see Supplementary Fig. S1 online). To check the structure, we carried out both powder and single-crystal x-ray diffraction experiments to confirm that $Pd_3P_2S_8$ forms in the space group of P-3m1 (see Fig. 1, and Supplementary Fig. S2, and Fig. S3 online)[41,42]. Afterward, we undertook the susceptibility measurement to find that it exhibits a sizeable diamagnetic response below the room temperature, as shown in Figure 2a and 2b[43] (see also Supplementary Fig. S4). The diamagnetic behavior is maintained down to the lowest measured temperature. There is a small upturn at much lower temperatures in the data, which is likely to come from the paramagnetic signals of the sample holder and adhesives used. Other than that, there is no hint of a phase transition down to 2 K, giving no reason to suspect any structural or magnetic phase transition over the temperature range of interest. The absence of structural transition can also be verified through heat capacity measurement (see Supplementary Fig. S2 online). It implies that the ideal kagome lattice is well protected over the temperature range of 2 to 300 K.

The material shows large diamagnetic signals due to the square-planar local structure around Pd ions. There is a small but apparent anisotropic behavior in the susceptibility. The susceptibility is $2.03 \pm 0.38 \times 10^{-4}$ emu/mol for the applied field perpendicular to the ab plane ($\chi_\perp$) and $1.75 \pm 0.52 \times 10^{-4}$ emu/mol for the parallel field ($\chi_\parallel$). Thus, the magnetic anisotropy of the sample is estimated to be about $\chi_\perp/\chi_\parallel = 1.18 \pm 0.59$. We note that this molar susceptibility is similar to bismuth and carbon structures, which exhibit large diamagnetic signals[44-46]. The diamagnetism of the material does not originate from the dynamics of the conducting materials. Instead, the Pd valence electrons with the planar local structure dominate the diamagnetic signals. The susceptibility is in accordance with the value calculated from the electronic band structure, as discussed later.

The optical transmittance measurements are carried out to inspect the electronic structures of $Pd_3P_2S_8$. From the transmittance measurements, we can extract the absorption coefficient and estimate the value of the fundamental bandgap using the Tauc plot method. From these data, we expect the fundamental bandgap to be about 2 eV, as shown in Figure 2c[47,48]. These basic characterizations of $Pd_3P_2S_8$ are consistent with the electronic band structures that will be presented below.

We examined the band structure of bulk and monolayer $Pd_3P_2S_8$ from first-principles calculations using a density functional theory (DFT) to find the unique electronic properties of this ideal 2D vdW kagome system. The effect of the vdW interactions is reflected in our calculated band structures by employing the experimentally measured lattice parameters. One should note that the vdW functionals used in DFT calculations are usually added to the total energy functional but do not alter the calculated electronic band structure. And the electrons are all paired, so the calculated electronic band structures have no spin labels. The calculated fundamental gap values are determined to be 1.78 and 1.40 eV for monolayer and bulk. The underestimation of the gap value due to the local approximations for the exchange-correlation interaction is estimated to be about 10 %. Regardless of the exchange-correlation functionals adopted, the energy bands obtained from the calculations are very flat, as shown in Figure 3 and Supplementary Figure S5 online. The flat band has a significant density of states (DOS) from the Pd $d$-orbitals of the 2D kagome lattice, whose energy is located right below the top valence bands. The Pd-dominant flat band over the whole reciprocal space remains to be robust for the accepted U value of ~1.5 eV for Pd atoms, which is a reasonable value according to previous studies.[49] As a further check of the validity of our calculated band structures, we were also able to extract the magnetic susceptibility using the obtained electronic band structures[50-52]. The calculated diamagnetic susceptibilities are $\chi_\perp = 2.38 \times 10^{-4}$ and $\chi_\parallel = 2.00 \times 10^{-4}$ emu/mol, respectively, when the field is perpendicular and parallel to the ab plane. The corresponding anisotropy ratio is $\chi_\perp/\chi_\parallel = 1.19$. These values are in good agreement with the experimental results.

For further details, the presence of kagome nature in bulk and monolayer $Pd_3P_2S_8$ can be assessed by examining the band touching at the high-symmetric k points. For an ideal kagome lattice, the flat band is expected to touch a dispersive band at the Γ point. This band touching at the Γ point is the crucial factor of the undisturbed 2D kagome system. The degeneracy at the Γ point of kagome systems is protected unless there are significant perturbations or disorders[22,23]. From the projected density of states (PDOS) at the Γ point of the top valence band (TVB), the dominant Pd *4d* character is confirmed. This band touching can be seen in the band structures of $Pd_3P_2S_8$, confirming the kagome nature.

Finally, the striking flatness of the top valence band (TVB) exists over a wide range of the reciprocal space, especially in the monolayer case. The extreme flatness of TVB can originate from two possible sources; one is a small electron hopping parameter between the atomic sites, and another is the CLS originated from a frustrated kagome structure. However, the large dispersion of the other bands touching the flat band at the Γ point does not support the small-hopping scenario. Thus, the extremely localized electron states representing the flat TVB suggest the presence of the unique nature of kagome lattice, which is the CLS. To rigorously confirm the CLS nature of the flat TVB, we analyzed the wavefunctions of the kagome model as obtained from our DFT calculation[53]. Figure 4 shows the contribution of Pd *d* orbitals at each site of the kagome lattice to the TVB and the phase of each orbital. Figure 4d plots the total contribution of each site to the flat TVB of DFT calculations (see also Supplementary Fig. S6 online). The contribution of each site is very similar to those of the ideal kagome model throughout the entire Brillouin zone of the Bloch wavevector (see the left column of Fig. 4). And the total contribution of atomic orbitals to the flat TVB is as substantial as 60 %. Our first-principles band calculations and the following model calculations of the ideal kagome structure are also in good agreement with each other regarding the phase difference among the orbitals forming the kagome lattice (see Fig. 4f to 4j). These results are strong evidence reinforcing our view that the flat TVB originates from the kagome structure.

Furthermore, considering that the Pd-dominant flat band is related to the diamagnetic response of the sample, the estimation of an 'effective' carrier effective mass over the entire Brillouin zone rather than at a specific wavevector were undertaken. We estimated the 'effective' mass, assuming a parabolic function for the energy in terms of the wavevector along each direction in momentum space. The 'effective' carrier mass of the bulk along G-K direction is 74.12 $m_e$, and 5.65 $m_e$ along G-A: where $m_e$ is the intrinsic mass of the free electron.

Moreover, we note that the flat region in the momentum space is wider for monolayer, as shown in Figure 3. The reduced flatness in the band structure of bulk $Pd_3P_2S_8$ can be understood similarly as mentioned in the recent work of $Fe_3Sn_2$[31]; the interlayer interaction is likely to interfere with the flat band nature and diminish the character. Therefore, it is natural to think that the difference between the projected density of states near the TVB of the bulk and monolayer $Pd_3P_2S_8$ is due to the interlayer interaction.

As an experimental demonstration, we produce monolayer by the mechanical exfoliation method, as shown in Figure 5. According to our experience, the cleavage energy of $Pd_3P_2S_8$ seems to be comparable to another S-based magnetic van-der-Waals materials of $NiPS_3$[9]. We note that the stability of the material was previously confirmed by theories like the phonon spectrum and ab-initio molecular dynamics calculation.[38] In this work, we have carried out extensive tests of stability under several conditions. According to our works, it is very stable in the air and shows no sign of degradation under several conditions for 25 hours. (see Supplementary Fig. S7 online) The availability of this monolayer sample opens up the possibility of examining the flat band physics in the 2D limit by gating experiments.

## Discussion

Materials having the kagome lattice have attracted significant interest in the field of condensed matter physics, in particular from the magnetism community, as it is believed to host an exotic spin liquid phase due to geometrical frustration. The electronic version of this magnetic frustration is the flat band in the electronic structure[54]. In this report, we have demonstrated the first ideal kagome vdW material, $Pd_3P_2S_8$, in which core ions are transition metal ions with strong spin-orbit coupling. By comprehensive experimental studies, we have confirmed the ideal kagome lattice remaining intact down to the lowest measured temperature and checked the fundamental physical properties that coincide with the theoretical calculations. By carrying out first-principles calculations, we have found that it also possesses an anomalously flat band at the top of the valence band, which is possible compact-localized states expected of the kagome lattice. The flat band with possible CLS can host strong electron correlations and gives rise to other novel phenomena. The flatness is more extensive in the form of a monolayer, verified by first-principle calculations. The kagome flat band is located near the Fermi level, making it very attractive to modulate the filling portion of the flat band on a real 2D limit with monolayer samples by gating experiments.[55,56] With the intriguing possibility of hosting topological states and emergent magnetism, $Pd_3P_2S_8$ will be proven to be an attractive platform to address some of the exciting physics related to kagome lattice on the 2D limit.

## Methods

### Crystal Synthesis

The synthesis of the crystal was done by the chemical vapor transport method with iodine as a transport agent. There were two methods reportedly used to synthesize the crystals. The first method was to use the atomic ratio of 3:2:8 for the Pd, P, and S, respectively. The ingredients (in total 1 g), iodine (0.05 g), and additional sulfur (0.05 g) were ground, mixed, and then sealed in a 20-cm-long quartz tube under a vacuum of lower $10^{-3}$ torr. Then the quartz tube was placed in a two-zone furnace. Zone 1 was heated to 720 °C for 6 hours and held there for 72 hours, while Zone 2 was kept at 690 °C. Subsequently, Zone 1 was cooled to 400 °C for 48 hours and maintained there for 6 hours; Zone 2 followed the same procedure with the target temperature of 350 °C. Finally, Zone 1(2) was cooled to 150 (120) °C for 6 hours before turning to natural cooling to room temperature. The second method is almost identical to the one previously used by Zhang *et al*[57].

## Structure Analysis

We carried out the structure analysis by measuring both powder and single-crystal samples using two diffractometers. A high-resolution powder diffraction experiment was performed using a commercial powder diffractometer, D8 discover (Bruker), with wavelengths $\lambda_{K\alpha}$ of 1.540590 and 1.544310 Å ($K_{\alpha 1}$ and $K_{\alpha 2}$, respectively) and XtaLAB P200 (Rigaku) with Mo source ($\lambda_{K\alpha}$ = 0.710747 Å) was used to identify the structure of the single-crystals. Both results support that our samples form in the space group of P-3m1.

## Magnetization Measurement

The magnetic susceptibility was measured with a commercial magnetometer (MPMS 3). About 2~10 mg of samples were used in each measurement and were attached to a quartz sample rod with GE varnish. For the thin samples, a slide cover glass was cut and used for the support. Diamagnetic signals of GE varnish were separately measured to be compared with the signals from the sample. For some measurement sets with the thin glass support, the background signals were subtracted off to obtain the pure sample signals and ensure that the significant signals were from the sample alone. The subtraction was done using the SquidLab software suite, which was developed to improve the fidelity of the signals during this project[43]. The magnetization was measured in two ways: the field sweeping at a specific temperature and the temperature sweeping with a particular applied field. The temperature sweepings were conducted with an applied field of 1 Tesla, and the field sweepings were conducted at several different temperatures of 2, 50, 100, 150, 200, 250, and 300 K.

## Optical Measurement

The optical measurements were carried out at 300 K on a grating spectrophotometer (Cary 5000, Agilent) over the spectral range of 190 to 2000 nm (0.62 to 6.5 eV) at a resolution of 1 nm with a spectral bandwidth of 2 nm. The absorption coefficient α was extracted from the transmittance T using Beer's law ($\alpha$ = - lnT/d) where d is the thickness. The single-crystal samples were thinned down by mechanical exfoliation and were fixed by using Kapton tape on Cu sample holders with an aperture of 0.5 mm diameter.

## Monolayer Exfoliation

The monolayer exfoliation was done onto a $SiO_2$/Si substrate. We used 285 nm thickness $SiO_2$ after comparing the optical visibility with other substrates[58]. The substrate was cut to approximately $1 \times 1$ cm$^2$ and was sonicated in isopropanol for 5 minutes. After blow-drying the substrate, it underwent oxygen plasma treatment with a flow rate of 100 sccm and 300 mT for 10 minutes to clean the surface[59]. Then the sample was exfoliated by one-sided tape before being placed onto the substrate. When necessary, the tape was very gently rubbed with a cotton swab after being attached to the substrate to eliminate any air bubbles and wrinkles. It was placed on the glass slide for annealing at 85 °C for 1 minute, after which the substrate was rapidly cooled with a small amount of acetone sprayed to the back of the glass. The tape was very slowly peeled off from the substrate leaving only a tiny tape residue behind.

## Theoretical Calculations

The electronic band structure and projected densities of states were calculated using DFT as implemented in the Quantum-ESPRESSO package[60,61]. Ion-electron interactions were described by the optimized norm-conserving Vanderbilt pseudopotentials[62,63], and the exchange-correlation energies were approximated using the Perdew, Burke, Ernzerhof functional[64]. We performed simulations on the monolayers by setting the interlayer distance to 15 Å. We used $8 \times 8 \times 8$ k-point mesh for the bulk crystal and a $12 \times 12 \times 1$ mesh for the monolayer and set the kinetic energy cutoff to 100 Ry. The atomic positions were optimized while fixing the lattice constants to the measured one[65].

## Data availability

*Scientific Reports* requires the inclusion of a data availability statement with all submitted manuscripts, as this journal requires authors to make available materials, data, and associated protocols to readers.

Figure legends (provided in numerical order)

Figure 1. a) Planar view of the lattice structure of $Pd_3P_2S_8$. The palladium ions form a two-dimensional kagome lattice. b) The kagome lattice is a geometrically frustrated system and has an inherent flat band structure. The flat band corresponds to a compact localized state known to host correlation effects that introduce topological states. c) Structure of $Pd_3P_2S_8$ viewed in the ac plane. The structure is slightly tilted to show the configuration of the 2D kagome structure within the plane. The monolayer of the material is illustrated.

Figure 2. a) Field sweep and b) temperature sweep measurement data are presented for the perpendicular and the parallel fields to the ab plane. The anisotropy of the diamagnetic susceptibility is displayed with the value of 1.18 on average. (◇ – parallel field; ● – perpendicular field) c) Bandgap estimated from the transmittance data. The estimated fundamental bandgap is about 2 eV using the Tauc plot method. (◇ – $(αhν)^{1/2}$; ○ – $(αhν)^2$) α is the absorption coefficient.

Figure 3. DFT calculation results for a) the monolayer and b) bulk crystal. Notably, the top valence band becomes flat as the crystal is exfoliated into monolayer, as shown in a), due to the weaker interlayer interactions. The projected density of states (PDOS) of palladium is dominant at the flat band.

Figure 4. The contributions of a) $d_{z^2}$ orbital, b) suitable linear combinations of $d_{xz}$ and $d_{yz}$ orbitals, and c) appropriate linear combinations of $d_{x^2-y^2}$ and $d_{xy}$ orbitals at each Pd site to the flat TVB. d) is the total contribution for all $d$ orbitals. A small dip at the M point indicates the hybridization between $d$ orbitals and lower band. Even in the presence of orbital hybridizations, as large as 60 % of the flat band comprises the atomic $d$ orbital of Pd. One can see that the projected contributions are very similar to those of e) ideal kagome model. Each atomic site is indicated by color. f-i) shows the phase of each orbital. Here, we set the phase of the $d_{z^2}$ orbital at Pd3 to zero. j) demonstrates the corresponding phases of the kagome model. The phase obtained from our first-principles calculations and those obtained from the model calculation are almost identical.

Figure 5. Optical microscope image and AFM scanned image of the flake on 285 nm $SiO_2$. Additional layers are also seen in the AFM image, forming steps. The yellow line crosses the boundary of the monolayer, and the height difference is about 1.25 nm, as indicated in the figure. The slight bump at the boundary indicates the tape residue on the sample.

Figure 1.

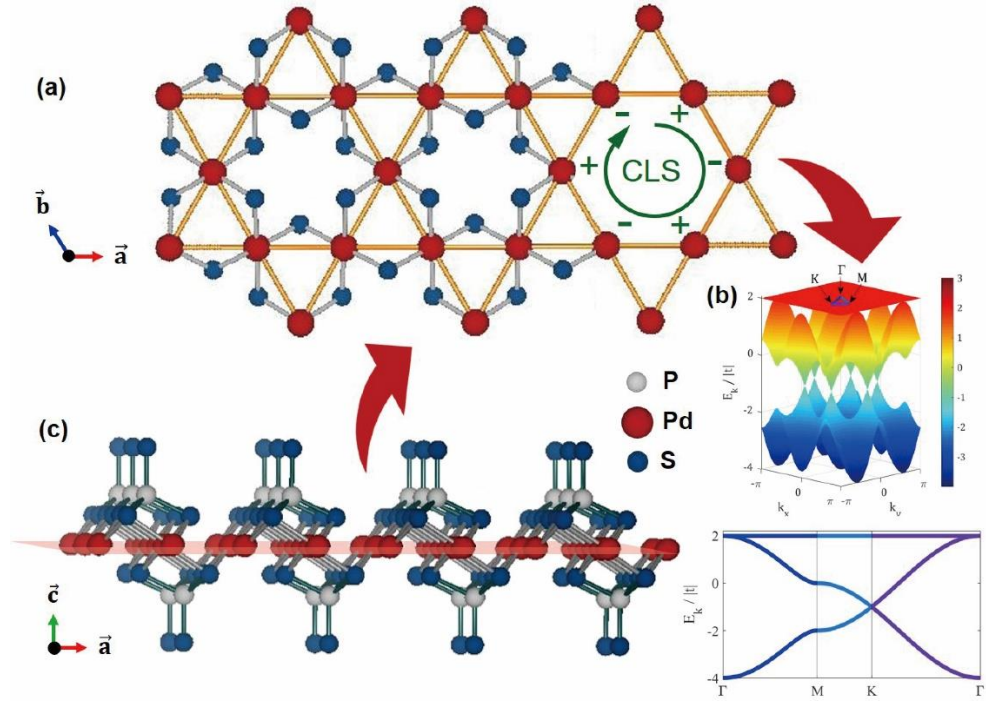

Figure 2.

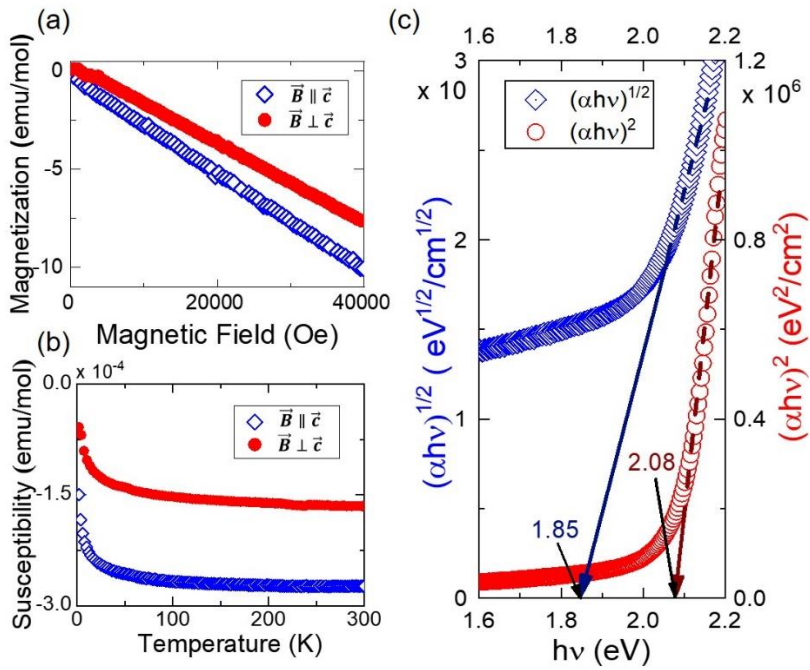

Figure 3.

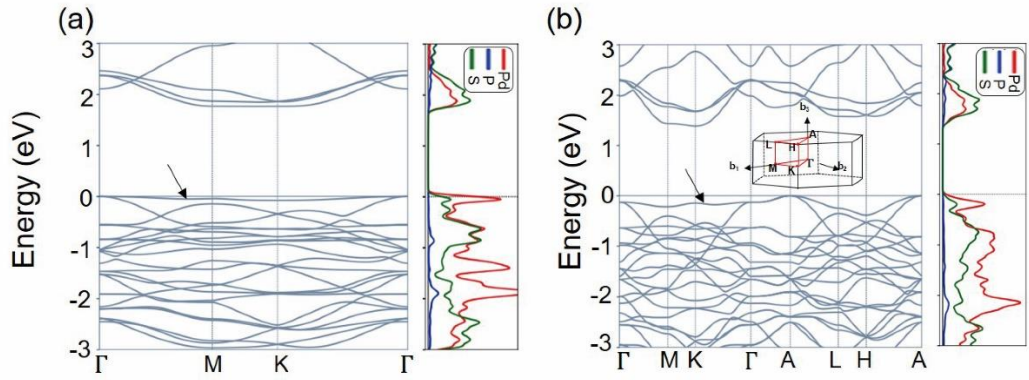

Figure 4.

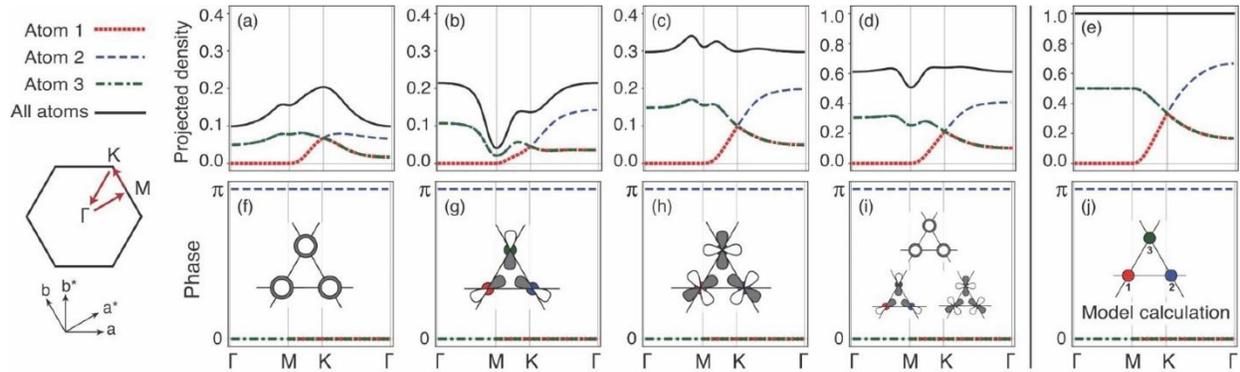

Figure 5.

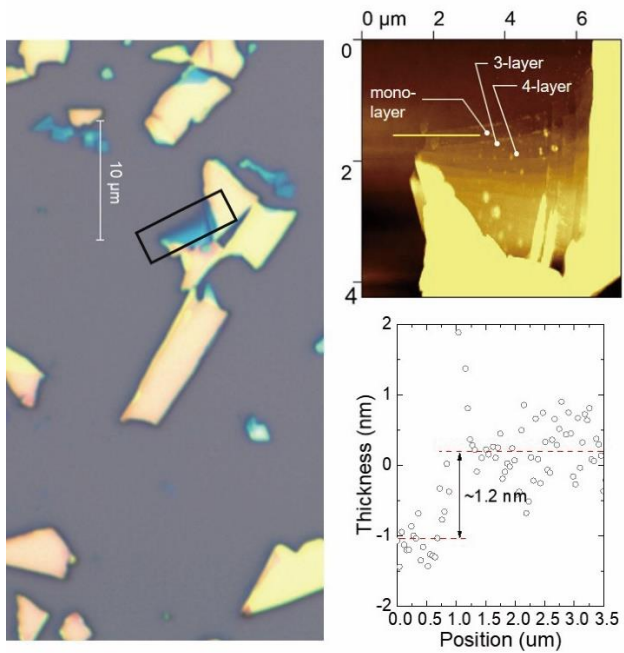

## Acknowledgments (optional)

We would like to acknowledge Maxim Mostovoy and Daniel Khomskii for helpful discussions. IBS-CCES was supported by the Institute for Basic Science (IBS) in Korea (Grant No. IBS-R009-G1). J. H. K. acknowledges support by the National Research Foundation of Korea (NRF) grants funded by the Ministry of Science and ICT (MSIT) of Korea (No. 2019R1I1A2A01062306, No. 2017R1A5A1014862: SRC program (vdWMRC)). Computational resources have been provided by KISTI (KSC-2018-CHA-0051). J.G.P. was supported by the Leading Researcher Program of the National Research Foundation of Korea (Grant No. 2020R1A3B2079375).

## Author contributions

J.-G.P. and S. P. conceived the experiments. S. P., S. K., and B. K. synthesized the materials. S. P., and S. K. carried out X-ray diffraction experiments and analyzed data together with Y. N. S. P., N. L., H. K. And S. K. performed the magnetization measurements while S. S. measured the heat capacity. M. J. C. and S. P. carried out the detailed data analysis. S. P. and J. (Junghyun) K. completed monolayer exfoliation. J. (Jonghyeon) K., K. I. S., and J. H. K. performed the optical measurements. P. K. and C. H. P. performed the DFT band calculations. K. L. carried out magnetization calculations and together with S. P. analyzed the experimental data. S. P., K. L., and J.-G. P. wrote the paper with contributions from all authors.

# Additional information

## Competing interests (mandatory)

The authors declare no competing interests.